# $^{75}$As-NMR Studies of LaFeAsO$_{1-x}$F$_x$ for Various $x$ Values

Yoshiaki Kobayashi[1,3]*, Erika Satomi[1], Sang Chul Lee[1], and Masatoshi Sato[2,1,3]

[1]*Department of Physics, Division of Material Science, Nagoya University, Furo-cho, Chikusa-ku, Nagoya 464-8602, Japan*

[2]*Toyota Physical & chemical Research Institute, Nagakute, Aichi 480-1192, Japan*

[3]*JST, TRIP, Nagoya University, Furo-cho, Chikusa-ku, Nagoya 464-8602, Japan*

Results of $^{75}$As NMR and NQR measurements are reported for superconducting samples of LaFeAsO$_{1-x}$F$_x$ with various $x$ values. The $x$ dependent widths of the superconducting transition and those of the NQR spectra indicate that the optimally doped LaFeAsO$_{1-x}$F$_x$ sample with the superconducting transition temperature $T_c$ ~28 K has the smallest width or smallest inhomogeneity of the superconducting order parameter $\Delta$. In the analyses of the temperature ($T$) dependence of the NMR longitudinal relaxation rate $1/T_1$, we have tried to use a simple relation $1/T_1 \sim T^n$ to see how fast $1/T_1$ decreases when the superconducting order parameter grows. The relation holds rather well in the $T$ region of ~$0.3T_c < T \leq T_c$, and $n$ varies from 2.5-6 with varying $x$, having the maximum value 5-6 at the optimal $x$. These results indicate that the $x$ dependence of $n$ originates from the spatial distribution of $\Delta$, and that the $T^{5-6}$ behavior is considered to be intrinsic to this system.

In the study of Fe pnictide superconductors,[1] to identify the symmetry of the superconducting order parameter $\Delta$ is important, because it is closely connected with the mechanism of the Cooper-pair formation. On this point, the so-called $S_{\pm}$ symmetry has been proposed as the most probable one,[2,3] and used in various arguments. It is characterized by the nodeless and opposite signs of $\Delta$ on each of two kinds of disconnected Fermi surfaces around $\Gamma$ and M points in the reciprocal space. To clarify whether they really have this symmetry, we have carried out various experimental studies.[4-10] For example, impurity effects on the superconducting transition temperature $T_c$, transport properties of RFeAsO$_{1-x}$F$_x$ (R1111) system (R=La and Nd)[4-7,9,10] and $^{75}$As NMR longitudinal relaxation rate $1/T_1$ of La1111 systems[8,10] have been studied. As the results, we have found that the $T_c$ suppression by the pair breaking expected for the $S_{\pm}$ symmetry due to the electron scattering by nonmagnetic impurities is too small to be explained. In the $^{75}$As $1/T_1$-$T$ curves, there is no coherence peak below $T_c$. Using a simple relation $1/T_1 \propto T^n$ to describe the curves in a rather wide region of ~$0.3T_c < T \leq T_c$, we obtain the $n$ values between 2.5–6 for LaFeAsO$_{1-x}$F$_x$, depending systematically on their $T_c$ values and the $1/T_1T$ values above $T_c$ (at 50 K, for example). (As shown later, it indicates the systematic $x$ dependence of $n$.) The maximum value of $n$ correspond to the optimum $x$ or maximum $T_c$ (~28 K),[8,10] indicating that the relation $1/T_1 \propto T^{2.5-3}$ observed by many groups[11-14] and considered as the favorable evidence for the $S_{\pm}$ symmetry,[15-18] may not be universal behavior of the system. Therefore, if we want to use the relation $1/T_1 \propto T^{2.5-3}$ in the symmetry arguments, we have to carry out further experimental studies. Even the nonexistence of the coherence peak cannot be directly considered to be the evidence for the $S_{\pm}$ symmetry, because effects of the electron-energy broadening near $T_c$ may wipe out the coherence peak. In ref. 10, to explain the electron number dependence of $n$ observed for LaFeAsO$_{1-x}$F$_x$, we have proposed its close connection with the spatial distribution of $T_c$ or $\Delta$, and here, on the basis of $^{75}$As-$1/T_1$ measurements in both the normal and superconducting states of LaFeAsO$_{1-x}$F$_x$ samples including those in the underdoped $x$ region, we show the explanation is really appropriate.

All the samples of LaFeAsO$_{1-x}$F$_x$ used in the present $^{75}$As NMR measurements were prepared in the $x$ region of $0.11 \leq x \leq 0.25$ by the same method[5-7] in the polycrystalline form. (Here we use the nominal $x$ values.) The X-ray powder patterns of the samples were taken with CuK$\alpha$ radiation at a step of 0.01° of the scattering angle 2θ, and the lattice parameters $a$ and $c$ were estimated. From the $T$ dependences of the superconducting diamagnetic moments and electrical resistivity $\rho$, two kinds of $T_c$ values of LaFeAsO$_{1-x}$F$_x$, $T_{c\chi}$ and $T_{c\rho}$, respectively, were also estimated. Details of the $T_c$-estimation can be found in refs. 5-7. Then, these $T_{c\chi}$ and $T_{c\rho}$ are plotted in Fig. 1(a) against $c$ instead of $x$. This plot enables us to avoid the nominal $x$ values, which may not necessarily align in the right order of the true $x$ values. (Note that the lattice parameter $c$ is linear in the true $x$ as reported in refs. 19 and 20, for example.) We find that the samples with $c$~8.72 Å are located near the optimally doped point, and that the $c$ values smaller and larger than ~8.72 Å, are in the overdoped and underdoped regions, respectively. Generally speaking, if filamentary superconductivity first sets in with decreasing $T$, the resistive transition temperature $T_{c\rho}$ is higher than $T_{c\chi}$ determined by magnetic measurements. It can explain the observation in the overdoped region. (The differences between $T_{c\chi}$ and $T_{c\rho}$ are significantly large for the samples with $c \leq 8.708$ Å. It is expected even for a constant amplitude of spatial fluctuation of $x$ or $c$, because $dT_c/dx$ or $dT_c/dc$ is large in the $c$ region far from the optimal $x$ point.) In the underdoped region, the difference between the two kinds of $T_c$ is rather small, and in contrast to the case of the overdoped region, the $T_{c\chi}$ seems to be slightly higher than $T_{c\rho}$. On this point, we discuss later in relation to the NMR profiles.

$^{75}$As-NMR and NQR measurements of LaFeAsO$_{1-x}$F$_x$ were carried out using the standard coherent pulse method for various samples labeled A-F [see Figs. 1(a) and 1(b)] from the overdoped to underdoped regions. (As stated above, the region $c$>~8.72 Å corresponds to the underdoped

*corresponding author (i45323a@cc.nagoya-u.ac.jp)

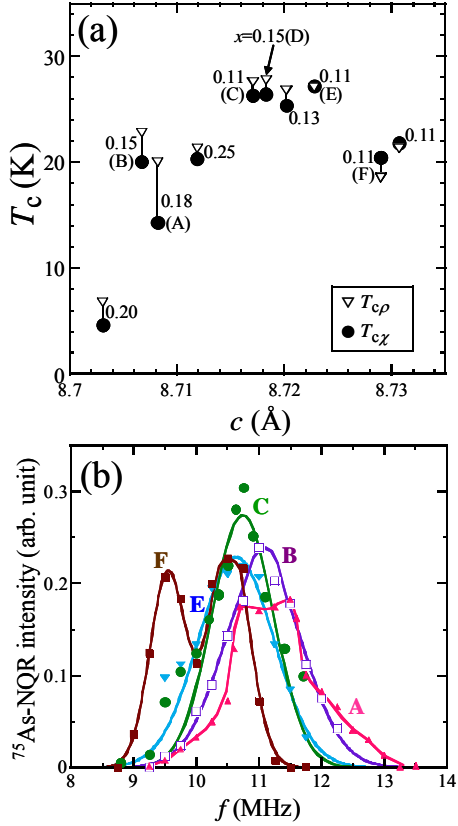

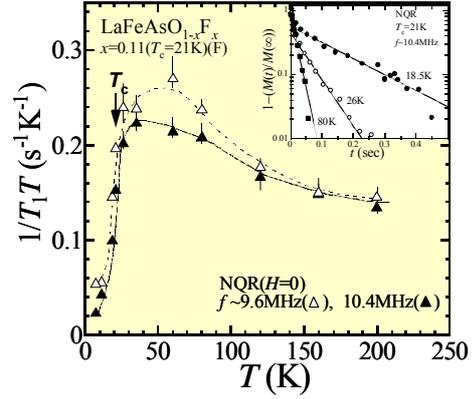

Fig. 1. (Color online) (a) Superconducting transition temperatures $T_c$ of LaFeAsO$_{1-x}$F$_x$ are plotted against the lattice parameter $c$. The data shown by inverted triangles (solid circles) were obtained from resistivity (superconducting diamagnetic) measurements. The detail of the estimation of $T_c$ is shown in ref. 6. Samples labeled by A-F are used in $^{75}$As NQR and NMR measurements. (b) $^{75}$As-NQR spectra of LaFeAsO$_{1-x}$F$_x$ are shown for various nominal $x$ values. The data A-F corresponds to the samples for which the $T_c$ values are

Fig. 2. (Color online) The nuclear relaxation rate $1/T_1$ divided by temperature $T$, $1/T_1T$ of the underdoped sample F is plotted against $T$. The $1/T_1T$ measured at the peak position of NQR spectra of ~9.6 (~10.4) MHz are shown by the open (closed) triangles. The inset shows the $1-M(t)/M(\infty)$-$t$ curves, where $M(t)$ is the nuclear magnetization at the time $t$ elapsed after irradiated the saturation RF-pulses.

region). The spectra were measured by recording the integrated spin-echo intensity $I$ with the frequency changed stepwise. The typical NMR spectra below 40 K can be found in refs. 6. The NQR spectra at 40 K are shown in Fig. 1(b). While optimally doped samples C and E have the almost single peaked NQR spectra, the NQR spectra of A (overdoped) and F (underdoped) seems to consist of more than two peaks and broader than those of C and E. Using these profiles, the $^{75}$As-NQR frequency $\nu_Q$ are determined at the peak position or at the mean value of the positions of the multiply-peaked $^{75}$As-NQR spectra. They have an increasing tendency with decreasing $c$ (or with increasing the true $x$). The $x$-dependences of the shapes and positions of the $^{75}$As-NQR spectra of LaFeAsO$_{1-x}$F$_x$ are similar to those reported previously by Lang et al.[21] The increasing width of the $^{75}$As-NQR spectra can be considered to reflect the increasing width of $x$, and this larger width of $x$ is expected to induce the larger widths of $T_c$ and $\Delta$, even though $T_c$ and $\Delta$ are smoothed over the spatial range of the coherence length. It can explain the observed large widths of $T_c$ in the overdoped region. (Note that the $x$ width is not so large as to be meaningfully detected by the X ray measurements of the Bragg reflection widths.)

For A-E, the $^{75}$As-NMR relaxation rates $1/T_1$ were estimated by using the recovery curves obtained from the integrated spin-echo intensity at the peak positions of the spectra corresponding to $\mathbf{H} \parallel ab$. For the underdoped sample F, $1/T_1$ was measured at the two peak positions observed in the NQR spectra (~9.6 and 10.4 MHz), and the $T$-dependence of $1/T_1T$ is shown in Fig. 2. The inset of the Fig. 2 shows the typical examples of the relaxation curves of the nuclear magnetization $M(t)$ against the time $t$ elapsed after irradiating the saturation RF pulse. The $1/T_1$-values were estimated by fitting the theoretical curves for the nuclear spin $I = 3/2$, $1 - M(t)/M(\infty) = f \times \exp(-3t/T_1)$ where $M(\infty)$ is the nuclear magnetization at thermal equilibrium. As shown in the inset, at the two peak positions of the spectra, the $f$-values are almost unity at high temperatures above ~50 K and they become smaller than unity below ~50 K possibly due to the incomplete saturation. Even this low-$T$ region, the $f$-values at the two peak positions are almost equal at given temperatures. We have also found that the peak positions, and the spectral shape and intensity do not exhibit appreciable change in the entire $T$ region studied here. The deviation of $f$ from unity observed below ~50 K does not indicate the spatial inhomogeneity of $1/T_1$, as discussed later. For the sample F, the $T$-dependence of $1/T_1T$ observed at the peak positions, $\nu_Q$ ~9.6 MHz, is very similar to that at another peak position, $\nu_Q$ of ~10.4MHz, and the magnitude of the former is slightly larger than that of the latter. Since the lower and higher $\nu_Q$ values (~9.6 and ~10.4 MHz respectively) are close to those of the antiferromagnetic[21] and optimally doped samples, respectively, it was first expected that the magnitudes and the $T$-dependence of $1/T_1T$ at the lower and higher $\nu_Q$ values were different. However, the $1/T_1T$ values observed for these two peaks exhibit almost the average of those expected at the peak positions, indicating that two parts having $\nu_Q$ ~9.6 and 10.4 MHz microscopically coexist.[21] Because this averaging behavior can be found even below $T_c$, the Cooper limit condition has to be satisfied, indicating that the characteristic length scale of the microscopic coexistence should be less than the superconducting coherence length. (The characteristic length of the nuclear spin diffusion is considered to be smaller than the superconducting coherence length.) This idea is consistent with the fact that we have not observed significant difference between $T_{c\chi}$ and $T_{c\rho}$ in the underdoped $x$ region. Because the $1/T_1$-measurements were performed under the condition with the nuclei only at one peak position in the NQR spectra being saturated, the energy transfer from the saturated nuclei to those at the other peak position may occurs. The effect of this energy transfer (spin diffusion) is observed as the rapidly recovering component of the spin-echo intensity in the $T$ region where the $1/T_1$ is small. It can explain the deviation of the $f$-value from unity below $\sim 50$ K.

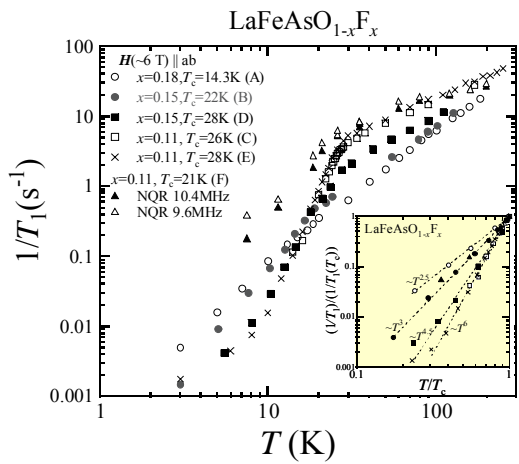

Fig. 3. (Color online) Main panel shows the $1/T_1$-$T$ curves of LaFeAsO$_{1-x}$F$_x$ plotted with the logalithmic scales for various nominal $x$ values. The samples A-E and F are measured by NMR ($H \sim 6$ T) and NQR (zero field), respectively. The inset shows the $1/T_1$-$T$ curves below $T_c$, where both axes are scaled by values at $T_c$. The power law behaviors down to about $0.2T_c$ can be observed. The optimally doped sample E has the largest $n$ value.

Figure 3 shows the $T$-dependence of $1/T_1$ with the logarithmic scales. They were obtained from the $^{75}$As-NMR data for the samples A-E at the field of $\sim 6$ T, and from the $^{75}$As-NQR data for the sample F. In the inset, $1/T_1$ scaled by the value at $T_c$ is plotted against $T/T_c$ in the same way. We find that $1/T_1$-values measured for the sample F at two $\nu_Q$ positions exhibit similar $T^3$-dependence below $T_c$. As we reported previously,[10] $n$ decreases from 5-6 to 2-3 with increasing $x$ from the optimal value. Now, we have also found that $n$ decreases, when the system shifts toward the underdoped side from the optimal point, or as $c$ increases from the optimal value of $\sim 8.72$ Å. This result indicates that the samples with the maximum $T_c$ have the maximum $n$, while the underdoped and overdoped samples have smaller values of $n$. Although $1/T_1$ deviates from the power law behavior in the region $T < \sim 0.3T_c$, it may be due to certain other nuclear relaxation processes: The one caused by vortex motions, for example, may become dominant in this $T$ region, where the contribution of the quasi-particles to $1/T_1$ is very small.

The observation of the large exponent ($n \sim 5$) of the relation $1/T_1 \propto T^n$ has also been reported by Hammerath et al.[22] for their La1111 sample having similar values of $T_c$ and lattice parameters to those of our optimally doped samples. They argue that the existence of the As-deficiency of 0.05–0.10 is essential for this large $n$-value. However, if the As-deficiency really exists, the doped electron number by the As deficiency has to be as much as 0.15-0.30, and the superconductivity will be suppressed: As reported in refs. 10 and 20, the disappearance of the superconductivity occurs when the hole-Fermi surface around the $\Gamma$ point diminishes upon the electron doping. We doubt whether their samples have "As-deficiency". The Rietveld analysis on our sample C indicates that the As-deficiency is less than 0.003. After these arguments, we can safely say that at the optimum $x$ value, the relation $1/T_1 \propto T^{5-6}$ can be considered to describe the intrinsic behavior of the optimum La1111 system.

We summarize, in Fig. 4(a), the data of $1/T_1T$ for all the samples A-F in linear scales. Note that in the high temperature region, the NQR-$1/T_1T$ of the underdoped sample F is smaller than the NMR-$1/T_1T$ of the optimally doped samples E observed under $H \sim 6$ T within the $ab$-plane. It might seem to contradict one of the well-known facts on the behavior of $1/T_1$ of the present system that, as $x$ changes from the underdoped to overdoped region through the optimally doped point, the $1/T_1T$-values above $T_c$ decreases.[11] We can explain this observation, considering the anisotropy of the hyperfine field from Fe sites. The anisotropy of $^{75}$As-$1/T_1$ for the Fe pnictides was discussed in refs. 23 and 24. In the present NQR measurements, the quantization axis of the nuclear spin is parallel to the $c$-axis, while in the NMR measurements the quantization axis of the nuclear spin is parallel to the $ab$-planes. Considering this difference, the ratio of the observed $1/T_1T$ values of the samples E and F can be estimated about 1.3 at $\sim 200$ K, which is close to the ratio $(1/T_1)_{ab}/(1/T_1)_c \sim 1.5$ obtained by the consideration of the isotropic spin fluctuation and the off diagonal components of the hyperfine field for the stripe type antiferromagnetic correlation in the parent compound.[23, 24]

In Fig. 4(b), the exponent $n$ is shown against $1/T_1T$ at 50K in the same form reported in ref. 10. [In this case, we use $1/T_1T$ as the abscissa instead of the nominal $x$, in a similar manner in Fig. 1(a) where the abscissa $c$ is used, because $1/T_1T$ can be considered, as stated above, to be a monotonic function of the true $x$-value. The use of $1/T_1T(50K)$ as the measure of $x$ is also rationalized by another well-known fact that the temperature derivative of $1/T_1T$ changes from negative to positive with decreasing $x$.] We clearly see that $n$ varies systematically with $1/T_1T(50$ K$)$ or $x$. Its maximum is located at around the optimally doped point, where the spatial distribution width of $\Delta$ is the smallest. On the basis of the observed data shown in Figs. 1(a) and 1(b), we think that the spatial inhomogeneity of $T_c$ and $\Delta$ is important for this phenomenon.[10]

Here, we also consider two other explanations of this $x$ dependent behavior of $n$. One is the $x$ dependent suppression of the magnetic fluctuation by the occurrence of the superconductivity,[25] and the other is the existence of two different superconducting gaps.[26]

The former possibility is considered for systems with the relatively strong spin fluctuations, in which $1/T_1T$ is enhanced by a factor of $\langle 1/(1-\alpha_q)^2 \rangle$,[27] where $1/(1-\alpha_q)$ is the $q$-dependent enhancement factor ($q$ is wave number of the spin fluctuations), and the symbol of $\langle \cdots \rangle$ denotes the averages on the Fermi surface. Because $\alpha_q$ decreases rapidly with decreasing $T$ due to the gap formation in the superconducting phase, $1/T_1T$ decreases more rapidly in systems with larger enhancement factor or stronger spin fluctuations. Then, because the spin fluctuation is stronger for the smaller $x$ as shown in Fig. 4(a), the faster suppression of $1/T_1T$ or the larger $n$ is expected. If this explanation works well, $n$ is expected to be larger for systems closer to the antiferromagnetic phase, which is not experimentally observed. One might think that the double peak structure observed in the underdoped sample F indicates that a certain phase separation exists, and that due to this phase separated nature, the $n$ value of the sample becomes smaller than that of the optimally doped ones. However, we can see that both NQR peaks of the sample are at the frequencies smaller than that of the optimally doped sample, which indicates that the $x$-values of both "phase separated" parts are smaller than the optimally doped sample. Therefore, the exponent $n$ of the sample F, which has, as stated above, the averaged value of those of the two parts, should be larger than that of the optimally doped sample. Thus, the suppression of the enhancement factor below $T_c$ seems not to explain the $x$ dependence of $n$.

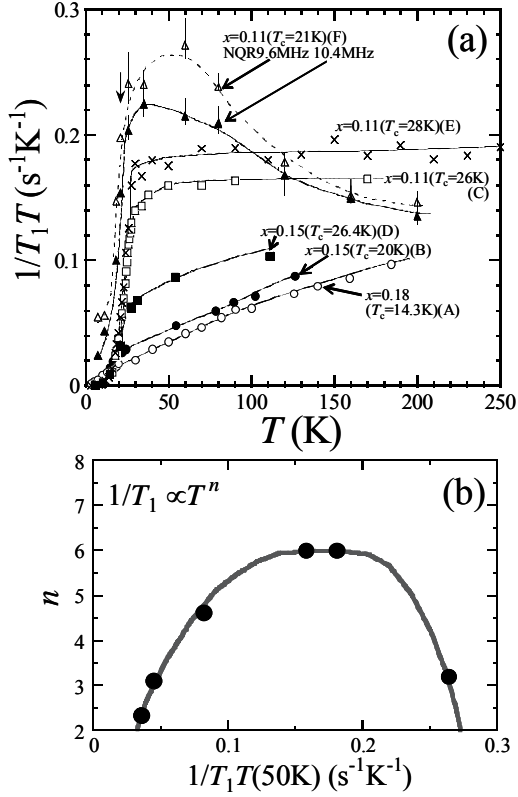

Fig. 4. (a) $1/T_1T$-$T$ curves obtained by the measurements of NMR (samples A-E; $H \parallel ab$) and NQR (sample F). (b) The $n$ values of the samples A-F plotted as a function of $1/T_1T$(50 K). Here, the $1/T_1T$(50 K) values are used as the measure of the $x$-value, because they change monotonically as a function of the true $x$-value.

The second explanation considers two gaps. From the angle-resolved photoemisison spectroscopy,[28] the gap $\Delta$ on the Fermi surface around the $\Gamma$ point is found to be smaller than that around the $M$ points. Therefore, when $x$ increases, the larger gap $\Delta_L$ on the electron Fermi surface becomes more important, and $n$ is expected to become larger with increasing $x$, which clearly contradicts the observation. Although the $T^3$ dependence of $1/T_1$ below $T_c$ has been explained by considering the existence of two different gap values,[13] it may not be able to explain the fact why different $n$ values were reported by many groups even for the samples with the highest $T_c$ values. (Here, we do not exclude the existence of different gap values.)

After all, to explain the behaviors of $n$, it is important to consider the spatial inhomogeneity of $\Delta$. On this view point, the $T^3$-like behaviors of $1/T_1$ below $T_c$ is considered not to be intrinsic, and the larger $n$-value of ~6 is intrinsic at least for La1111 system.

In summary, $^{75}$As NMR and NQR studies have been carried out for superconducting samples of LaFeAsO$_{1-x}$F$_x$ with various $x$ values. We have found that, in the relation $1/T_1 \propto T^n$ simply introduced to describe the $T$ dependence of $1/T_1$, $n$ has the maximum value at the optimal $x$. The exponent $n$ becomes smaller with increasing the spatial inhomogeneity of $\Delta$. In this sense, the $n$ value at the optimally doped samples is considered to be intrinsic.

**Acknowledgment**

This work is supported by Grants-in-Aid for Scientific Research from the Japan Society for the Promotion of Science (JSPS), and Technology and JST, TRIP.

1) Y. Kamihara, *et al.*: J. Am. Chem. Soc. **130** (2008) 3296.
2) I. I. Mazin, *et al.*: Phys. Rev. Lett. **101** (2008) 057003.
3) K. Kuroki, *et al.*: Phys. Rev. Lett. **101** (2008) 087004.
4) A. Kawabata, *et al.*: Int. Symp. Fe-Pnictide Superconductors, Tokyo, 2008.
5) A. Kawabata, *et al.*: *Proc. Int. Symp. Fe-Pnictide Superconductors,* J. Phys. Soc. Jpn. **77** (2008) Suppl. C, p. 147.
6) A. Kawabata, *et al.*: J. Phys. Soc. Jpn. **77** (2008) 103704.
7) S. C. Lee, *et al.*: J. Phys. Soc. Jpn. **78** (2009) 043703.
8) Y. Kobayashi, *et al.*: J. Phys. Soc. Jpn. **78** (2009) 073704.
9) S. C. Lee, *et al.*: J. Phys. Soc. Jpn. **79** (2010) 023702.
10) M. Sato, *et al.*: J. Phys. Soc. Jpn. **79** (2010) 014710.
11) Y. Nakai, *et al.*: J. Phys. Soc. Jpn. **77** (2008) 073701.
12) H.-J. Grafe, *et al.*: Phys. Rev. Lett. **101** (2008) 047003.
13) H. Mukuda, *et al.*: J. Phys. Soc. Jpn. **77** (2008) 093704.
14) S. Kawasaki, *et al.*: Phys. Rev. B **78** (2008) 220506.
15) A. V. Chubukov, *et al.*: Phys. Rev. B **78** (2008) 134512.
16) Y. Bang, and H.-Y. Choi: Phys. Rev. B **78** (2008) 134523.
17) D. Parker, *et al.*: Phys. Rev. B **78** (2008) 134524.
18) Y. Nagai, *et al.*: New J. Phys. **10** (2008) 103026.
19) S. W. Kim, *et al.*: *Proc. Int. Symp. Fe-Pnictide Superconductors,* J. Phys. Soc. Jpn. **77** (2008) Suppl. C, p. 23.
20) E. Satomi, *et al.*: J. Phys. Soc. Jpn. **79** (2010) 094702.
21) G. Lang, *et al.*: Phys. Rev. Lett. **104** (2010) 097001.
22) F. Hammerath, *et al.*: Phys. Rev. B **81** (2010) 140504.
23) K. Kitagawa, *et al.*: J. Phys. Soc. Jpn. **78** (2009) 063706.
24) K. Matano, *et al.*: Europhys. Lett. **87** (2009) 27012.
25) T. Kariyado, M. Ogata: arXiv:0911.2963; to be published in Physica C [DOI: 10.1016/j.physc.2009.10.123].
26) M. Yashima, *et al.*: J. Phys. Soc. Jpn. **78** (2009) 103702.
27) T. Moriya: J. Phys. Soc. Jpn. **18** (1963) 516.
28) T. Shimojima *et al.*: 65th Annu. Meet. Physical Society of Japan, March 2010.